%% file: main.tex
\documentclass{article}
\usepackage[letterpaper, left=1.3in, right=1.3in, bottom=1.25in, top=1.25in]{geometry}
\usepackage[utf8]{inputenc}
\usepackage[english]{babel}
\usepackage{comment}
\usepackage{authblk}
\usepackage{graphicx}
\graphicspath{{figures/}{../figures/}}
\usepackage{blindtext}
\usepackage{subfiles}
\usepackage{amsmath}
\usepackage{amssymb}
\usepackage{siunitx}
\usepackage{booktabs} 
\usepackage{hyperref}
\usepackage{float}
\usepackage{textcomp}
\usepackage[sort,comma,authoryear,round]{natbib}
\usepackage{tabularx}
\usepackage{chngcntr}
\usepackage{fancyhdr}
\usepackage[font=small,skip=0pt]{caption}
\usepackage[titletoc]{appendix}
\usepackage{multirow}
\usepackage{adjustbox}

\title{Comparing interpretation methods in mental state decoding analyses with deep learning models}

\author[$\cap$,$\vee$,$\circ$]{\small Armin W. Thomas}
\author[$\sqcup$]{\small Christopher Ré}
\author[$\cap$,$\vee$,$\circ$]{\small Russell A. Poldrack}

\affil[$\cap$]{\footnotesize Stanford Data Science, Stanford University, Stanford, CA, USA}
\affil[$\vee$]{\footnotesize Dept. of Psychology, Stanford University, Stanford, CA, USA}
\affil[$\sqcup$]{\footnotesize Dept. of Computer Science, Stanford University, Stanford, CA, USA}
\affil[$\circ$]{\footnotesize \{athms,russpold\}@stanford.edu}

\date{July 2022}

\begin{document}
\maketitle

\begin{abstract}
    Deep learning (DL) models find increasing application in mental state decoding, where researchers seek to understand the mapping between mental states (e.g., perceiving fear or joy) and brain activity by identifying those brain regions (and networks) whose activity allows to accurately identify (i.e., decode) these states. Once a DL model has been trained to accurately decode a set of mental states, neuroimaging researchers often make use of interpretation methods from explainable artificial intelligence research to understand the model's learned mappings between mental states and brain activity. Here, we compare the explanation performance of prominent interpretation methods in a mental state decoding analysis of three functional Magnetic Resonance Imaging (fMRI) datasets. Our findings demonstrate a gradient between two key characteristics of an explanation in mental state decoding, namely, its biological plausibility and  faithfulness: interpretation methods with high explanation faithfulness, which capture the model's decision process well, generally provide explanations that are biologically less plausible than the explanations of interpretation methods with less explanation faithfulness. Based on this finding, we provide specific recommendations for the application of interpretation methods in mental state decoding.
\end{abstract}

\subfile{sections/introduction}

\section{Methods}
\subfile{sections/methods/data}
\subfile{sections/methods/trial-spm}
\subfile{sections/methods/model}
\subfile{sections/methods/xai-methods}
\subfile{sections/methods/statistical-modelling}
\subfile{sections/methods/code-data-availability}

\section{Results}
\subfile{sections/results/hyperopt}
\subfile{sections/results/decoding-performance}
\subfile{sections/results/brain-maps}
\subfile{sections/results/faithfulness}
\subfile{sections/results/sanity-checks}

\section{Discussion}
\subfile{sections/discussion}


\paragraph{Acknowledgments.}
\subfile{sections/acknowledgments}

\bibliographystyle{apalike}
\bibliography{references,fmriprep_references}

\newpage
\appendix
\renewcommand{\thefigure}{A\arabic{figure}}
\setcounter{figure}{0}

\section{Methods}
\subsection{Data}
\subfile{supplements/methods/fmriprep}

\newpage
\section{Results}
\subfile{supplements/results/randomized-labels-performance}

\end{document}

%% file: sections/introduction.tex
Deep learning (DL) models have celebrated immense successes in many areas of research and industry ~\citep{lecun_deep_2015,goodfellow_deep_2016}. This success is often attributed to their unmatched ability to learn versatile representations of complex datasets, allowing them to associate a target signal with varying patterns in minimally-preprocessed (or raw) data. Due to their empirical success, neuroimaging researchers have started applying DL models to mental state decoding analyses~\citep[e.g.,][]{wang_decoding_2020,zhang_functional_2021,thomas_analyzing_2019,mensch_extracting_2021,vanrullen_reconstructing_2019,plis_deep_2014}. In these analyses, researchers seek to understand how specific mental states (e.g., answering questions about a prose story or math problem) are represented in the brain by identifying brain regions (or networks of brain regions) whose activity patterns allow accurate identification (i.e., decoding) of these mental states ~\citep{haynes_decoding_2006}.

Once a DL model has been trained to accurately decode a set of mental states from brain activity, researchers often make use of interpretation methods from explainable artificial intelligence research~\citep[XAI;][]{samek_explaining_2021,doshi-velez_towards_2017} to gain insights into the models' learned mappings between mental states and brain activity, seeking to overcome the un-interpretability of DL models~\citep{rudin_stop_2019}. From the wealth of existing interpretation methods~\citep{linardatos_explainable_2021,gilpin_explaining_2018}, neuroimaging researchers most often utilize attribution (i.e., heatmapping) methods, which attribute a relevance to each feature value of an input for the resulting decoding decision of a DL model, resulting in a heatmap of relevance values~\citep{samek_explaining_2021}. On a high level, prominent attribution methods in the mental state decoding literature can be grouped into sensitivity analyses~\citep[e.g.,][]{simonyan_deep_2014,springenberg_striving_2015,smilkov_smoothgrad_2017}, reference-based attributions~\citep[e.g.,][]{sundararajan_axiomatic_2017,shrikumar_learning_2017,lundberg_unified_2017}, and backward decompositions~\citep[e.g.,][]{bach_pixel-wise_2015,montavon_explaining_2017}. Sensitivity analyses, such as Gradient Analysis~\citep{simonyan_deep_2014}, attribute relevance to input features according to how sensitively a model's decoding decision responds to a feature's value. Reference-based attributions, such as DeepLift~\citep{shrikumar_learning_2017} or Integrated Gradients~\citep{sundararajan_axiomatic_2017}, by contrast, attribute relevance to input features by comparing the model's response to a given input to its response to a reference input (e.g., a neutral input). Backward decompositions, such as Layer-wise relevance propagation~\citep[LRP;][]{bach_pixel-wise_2015}, on the other hand, attribute relevance to input features by decomposing the decoding decision of a DL model in a backward pass through the model into the contributions of lower-level model units to the decision, up to the input space, where a contribution for each input feature can be defined. 

Given the wealth of existing attribution methods, neuroimaging researchers interested in interpreting the mental state decoding decisions of DL models are faced with the task of choosing a method for their particular analysis and research question. Yet, in many cases, the explanations of different attribution methods are difficult to visually discern, making it challenging to compare and evaluate their quality. Even further, it is unclear whether related empirical findings from computer vision~\citep[CV;][]{kindermans_reliability_2019,adebayo_sanity_2018,samek_evaluating_2017} and natural language processing~\citep[NLP;][]{jain_attention_2019,jacovi_towards_2020,ding_evaluating_2021}, on the relative performance of different attribution methods, generalize to neuroimaging data. There, researchers have often argued that reference-based attributions and backward decompositions are superior to sensitivity analyses because they capture the decision process of DL models more faithfully. Yet, mental state decoding is distinct from most CV and NLP applications in that researchers seek to understand the association of input data (i.e., brain activity) and decoding targets (i.e., mental states), whereas CV and NLP are often solely concerned with predictive performances~\citep{lipton_troubling_2018}. To date, it is therefore unclear how prominent attribution methods compare in providing insights into the association of brain activity and mental states learned by DL models. 

In this work, we compare the explanation performance of prominent attribution methods in a mental state decoding analysis of three functional Magnetic Resonance (fMRI) datasets. To compare performances, we use two main criteria: First, we evaluate the biological plausibility of the explanations of an attribution method by testing whether they identify all voxels of the input whose activity pattern is reliably associated with the decoded mental state. To this end, we compare its explanations to the results of a standard general linear model~\citep[GLM;][]{holmes_generalisability_1998} analysis of the fMRI data. We find that the explanations of sensitivity analyses are generally more similar to the results of a GLM analysis when compared to the explanations of reference-based attributions and backward decompositions. Second, to understand how well the explanations capture the decision process of the DL model, we evaluate their faithfulness by testing whether they correctly identify those voxels of the input whose activity is necessary for the model to accurately decode the mental states. We find that the explanations of reference-based attributions and backward decompositions are generally more faithful than those of sensitivity analyses.

Taken together, these findings lead us to a twofold recommendation for attribution methods in mental state decoding: If researchers want to understand the decision process of a DL model in mental state decoding, we recommend reference-based attribution methods (such as DeepLift~\citep{shrikumar_learning_2017}, DeepLift SHAP~\cite{lundberg_unified_2017}, and Integrated Gradients~\citep{sundararajan_axiomatic_2017}), and backward decompositions (such as LRP~\citep{bach_pixel-wise_2015}) because their explanations are the most faithful in our analyses. By contrast, if researchers want to understand the association between mental states and brain activity, and merely use DL models as a tool to study this association, we recommend sensitivity analyses (such as Gradient Analysis~\citep{simonyan_deep_2014}, SmoothGrad~\citep{smilkov_smoothgrad_2017}, Guided Backpropagation~\citep{springenberg_striving_2015}, and Guided GradCam~\citep{selvaraju_grad-cam_2017}) because their explanations align better with the results of standard analysis approaches for fMRI data.

%% file: sections/methods/data.tex
\subsection{Data}
\label{methods_data}
We analyzed three fMRI datasets in this study, namely, 
fMRI data of 44 randomly-selected individuals in the motor task of the 
Human Connectome Project~\citep[HCP;][]{van_essen_wu-minn_2013}, 44 
randomly-selected individuals in the HCP's working memory (WM) task, 
and 58 individuals in a pain and social rejection 
experiment published by~\citet{woo_separate_2014}. We refer to these 
three datasets respectively as "MOTOR", "WM", and "heat-rejection" in the following
and provide a brief overview of their experiment tasks as well as details 
on the fMRI acquisition and preprocessing. For any further methodological 
details, we refer the reader to the original publications~\citep{van_essen_wu-minn_2013,woo_separate_2014}.

\subsubsection{Experiment tasks}
\label{methods_experiment_tasks}

\paragraph{Heat-rejection: } The heat-rejection dataset comprises fMRI data from two
experimental tasks. In the rejection task, individuals either see head shots of an
ex-partner with a cue-phrase beneath the photo directing them to think about how 
they felt during the break-up (rejection) or a head shot of a close friend with a cue-phrase 
directing them to think about a specific positive experience with this friend 
(no rejection). In the somatic pain task, individuals focus on a hot (painful) or warm (not painful)
stimulus that is delivered to their left forearm (with temperatures calibrated
to each participant). Each rejection trial begins with a 7 s fixation cross, followed 
by a 15 s presentation period of a photo (ex-partner or friend),
a 5 s five-point affect rating period, and 18 s of a visuo-spatial control task in
which individuals see an arrow pointing left or right and are asked to indicate in
which direction the arrow is pointing. Heat trials are identical in structure to
rejection trials with the exception that individuals see a fixation cross 
during the 15 s thermal stimulation period (consisting of a 1.5 s temperature ramp
up/down and 12 s at peak temperature) and subsequently use the five-point rating 
scale to report their experienced pain level.

\paragraph{MOTOR:} In the HCP's motor task, individuals see visual cues asking them to 
tap their left or right fingers, squeeze their left or right toes or move their
tongue. The task was presented in blocks of 12 s, each including one
movement type, preceded by 3 s cue. Two fMRI runs were collected for this task, each
comprising two blocks of tongue movements, four blocks of hand movements (two
left, two right), and four blocks of foot movements (again, two left and two
right) as well as three 15 s fixation blocks.

\paragraph{WM:} In the HCP's WM task, individuals see images of one of
four different stimulus types (namely, body parts, faces, places or tools). In one half of the task
blocks, individuals are asked to indicate whether the current stimulus is the same 
as the stimulus that was shown 2 before (2-back). In the other half of the task blocks, 
individuals are asked to indicate whether the currently presented stimulus is the same
as a target stimulus that was shown at the beginning of the block (0-back). Two fMRI runs 
were collected for this task, each comprising eight task (25 s each) and
four fixation blocks (15 s each). Each task block consists of 10 trials (2.5 s each) of 
2 s stimulus presentation and 0.5 s interstimulus interval. Note that we pool the data of the two N-back conditions because we are not interested in identifying any effect of the N-back condition on brain activity.

\subsubsection{FMRI data acquisition}
\label{methods_fmri_acquisition}

\paragraph{Heat-rejection: }  Whole-brain EPI acquisitions were acquired on 
a GE 1.5 T scanner using a T2*-weighted spiral in-out sequence developed by Dr Gary Glover with TR = 2,000 ms, TE = 40 ms, flip angle = 84, FOV = 22 cm, and 
24 axial slices with $3.5 \times 3.5 \times 4.5$ mm voxels parallel to the anterior commissure-posterior commissure line~\citep[for further methodological details on fMRI data acquisition, see][]{woo_separate_2014}).

\paragraph{Human Connectome Project:} Whole-brain EPI acquisitions were acquired 
with a 32-channel head coil on a modified 3T Siemens Skyra with TR = 720 ms, 
TE = 33.1 ms, flip angle = 52, in-plane FOV = $20,8 \times 18 $cm, and 72 slices with 2.0 mm isotropic
voxels. Two fMRI runs were acquired for each task, one with a right-to-left and the other 
with a left-to-right phase encoding~\citep[for further methodological details on fMRI data acquisition, see][]{ugurbil_pushing_2013}.

\subsubsection{FMRI data preprocessing}
\label{methods_fmri_preprocessing}

\paragraph{Human Connectome Project:} We preprocessed the fMRI data of the MOTOR 
and WM datasets with fmriprep 20.2.3~\citep{esteban_fmriprep_2019}.
A detailed description of all preprocessing steps can be found in 
Appendix \ref{appendix_fmriprep_details}. 

\paragraph{Heat-rejection:} The data preprocessing was performed by the original authors ~\citep[see][]{woo_separate_2014} and included removal of the
first four volumes of each fMRI run to allow for image intensity stabilization,
slice timing correction (realignment) with SPM8, spatial warping to SPM’s normative 
atlas using warping parameters estimated from co-registered, high-resolution 
structural images, interpolated to $2 \times 2 \times 2$ mm voxels, and spatial smoothing with 
an 8 mm FWHM Gaussian kernel. 

%% file: sections/methods/trial-spm.tex
\subsection{Statistical parametric maps}

\subsubsection{Trial-level maps}
\label{methods_trial_glm}
We performed all of our analyses on trial-level voxel-wise statistical parametric maps~\citep{friston_statistical_1994} that were computed for each experiment 
trial of a dataset. We refer to these maps as trial-level blood-oxygen-level-dependent (BOLD) maps throughout the rest of the manuscript.

\paragraph{Heat-rejection: } Trial-level BOLD maps were computed by the original authors~\citep[see][]{woo_separate_2014} in an analysis that included boxcar regressors, convolved with the canonical haemodynamic response function (HRF), for the 15 s photo or pain periods, the subsequent 5 s affect or pain rating period, and the 18 s period of the visuospatial control task (leaving the fixation-cross periods as unmodeled baselines), and a boxcar regressor for each individual trial.  In addition, nuisance covariates of no interest were included in the analysis representing a linear drift across time within each run, the six estimated head movement parameters for each run (x, y, z, roll, pitch and yaw; mean-centered) as well as their squares, derivatives, and squared derivatives, and indicator vectors for outlier time points~\citep[for details on outlier detection, see ][]{woo_separate_2014}.

\paragraph{Human Connectome Project: } We computed trial-level BOLD maps by the use of Nilearn 0.9.0~\citep{abraham_machine_2014}. This analysis included boxcar regressors for each trial type (i.e., body, face, place, tool for the WM task and left/right foot, left/right finger, and tongue for the MOTOR task), which we convolved with a standard Glover HRF (as implemented by Nilearn~\citep{abraham_machine_2014}; leaving the fixation periods as unmodelled baselines), and a boxcar regressor for each individual trial. In addition, the analysis included nuissance regressors of no interest representing the six estimated head movement parameters (x, y, z, roll, pitch and yaw) as well as their squares, derivatives, and squared  derivatives, the average signal of white matter and cerebrospinal fluid masks, the  global signal, and a set of low-frequency regressors to account for slow signal drifts below 128 s.

\subsubsection{Subject- and group-level maps}
\label{methods_subject_group_glm}

To aggregate the trial-level BOLD maps to the subject- and group-level, we used a standard two-stage analysis procedure as proposed by~\citep{holmes_generalisability_1998}.

The subject-level analysis included a binary indicator variable for each mental state of a dataset, which we used to contrast each mental state of a dataset against all other mental states of the dataset. Note that the subject-level analysis of the two HCP datasets (WM and MOTOR) also included a binary nuissance variable for each of the two fMRI runs.

The group-level analysis included a binary indicator variable for each subject-level contrast type (i.e., mental state) as well as a binary nuisance variable for each included individual. Accordingly, the resulting group-level contrast maps correspond to a paired, two-sample t-test over the subject-level contrast maps. Note that we smoothed all subject-level contrast maps with a 5 mm FWHM Gaussian kernel in the group-level analysis.

\subsection{Training and test data}
\label{methods_data_split}
To create distinct training and test datasets, we separated the trial-level BOLD maps of each dataset by assigning the maps of every 5th individual of a dataset to a test dataset and designating the maps of all remaining individuals as training data.

%% file: sections/methods/model.tex
\subsection{Deep learning model}
\label{methods_model}

We use 3D-convolutional neural network architectures
~\citep[3D-CNNs; ][]{lecun_convolutional_1998} as mental state decoding models, which are composed of a stack of 3D-convolution layers and a dense output layer.

A 3D-convolution layer consists of a set of 3D-kernels that each learn specific 
features of an input volume $x$. Each kernel $k$ learns a volumetric feature that is
convolved over the input, resulting in an activation map $A^k$ that indicates the presence of the learned feature for each spatial location of the input:
$A_{i,j,l}^k = g(\sum_{m} \sum_{n} \sum_{z} w_{m,n,z}^kx_{i+m-1, j+n-1, l+z-1} + b^k)$. 
Here, $b^k$ and $w^k$ represent the bias and weights of the kernel, while $g$ represents the rectified 
linear unit activation function ($g(x) = max(0, x)$). The indices $m$, $n$, 
and $z$ indicate the row, column, and height of the 3D-convolution kernel, while $i$, $j$, 
and $l$ indicate the coronal (i.e., row), saggital (i.e., column), and axial 
(i.e., height) dimension of the activation map $A^k$. Note that the models move all convolution kernels over their input at a stride size of 2, thus applying the kernels to every other value of a layer's input. The models further apply batch-normalization  \cite{ioffe_batch_2015} to the linear outputs of each convolution layer (before the non-linear activation).

To make a decoding decision, the models flatten the activation maps 
$A$ resulting from the last convolution layer and pass the flattened maps to a dense softmax output layer, which predicts a probability $p_c$ that input $x$ 
represents mental state $c$: $p_c = \sigma(b_c + \sum_i a_i w_{ic})$, 
where $b$ and $w$ represent the layer's bias and weights, while 
$\sigma$ indicates the softmax function: $\sigma(x)_j = \frac{e^{x_j}}{\sum_i e^{x_i}}$.

\subsubsection{Training details}
\label{methods_training_details}
If not reported otherwise, we train models with stochastic gradient descent and the ADAM optimizer~\citep{kingma_adam_2017} for 40 training epochs, where one epoch is defined as an entire iteration over a dataset's training data. 

\subsubsection{Hyper-parameter evaluation}
\label{methods_hyperopt}
For each dataset, we perform a grid search to determine a set of best-performing model and optimization hyper-parameters. Specifically, we search over the number of model convolutional layers ($3$, $4$, and $5$), the number of kernels per layer ($4$, $8$, and $16$), and the kernels' size ($3$ and  $5$) as well as the mini-batch size ($32$ and $64$), learning rate ($1e^{-4}$ and $1e^{-3}$), and dropout rate ($25\%$ and $50\%$; applied to convolution layers) used during training (for an overview of the grid search results, see section \ref{results_hyperopt}).

%% file: sections/methods/xai-methods.tex
\subsection{Attribution methods}
\label{methods_xai_approaches}

\begin{figure}[!t]
\begin{center}
\includegraphics[width=0.9\linewidth]{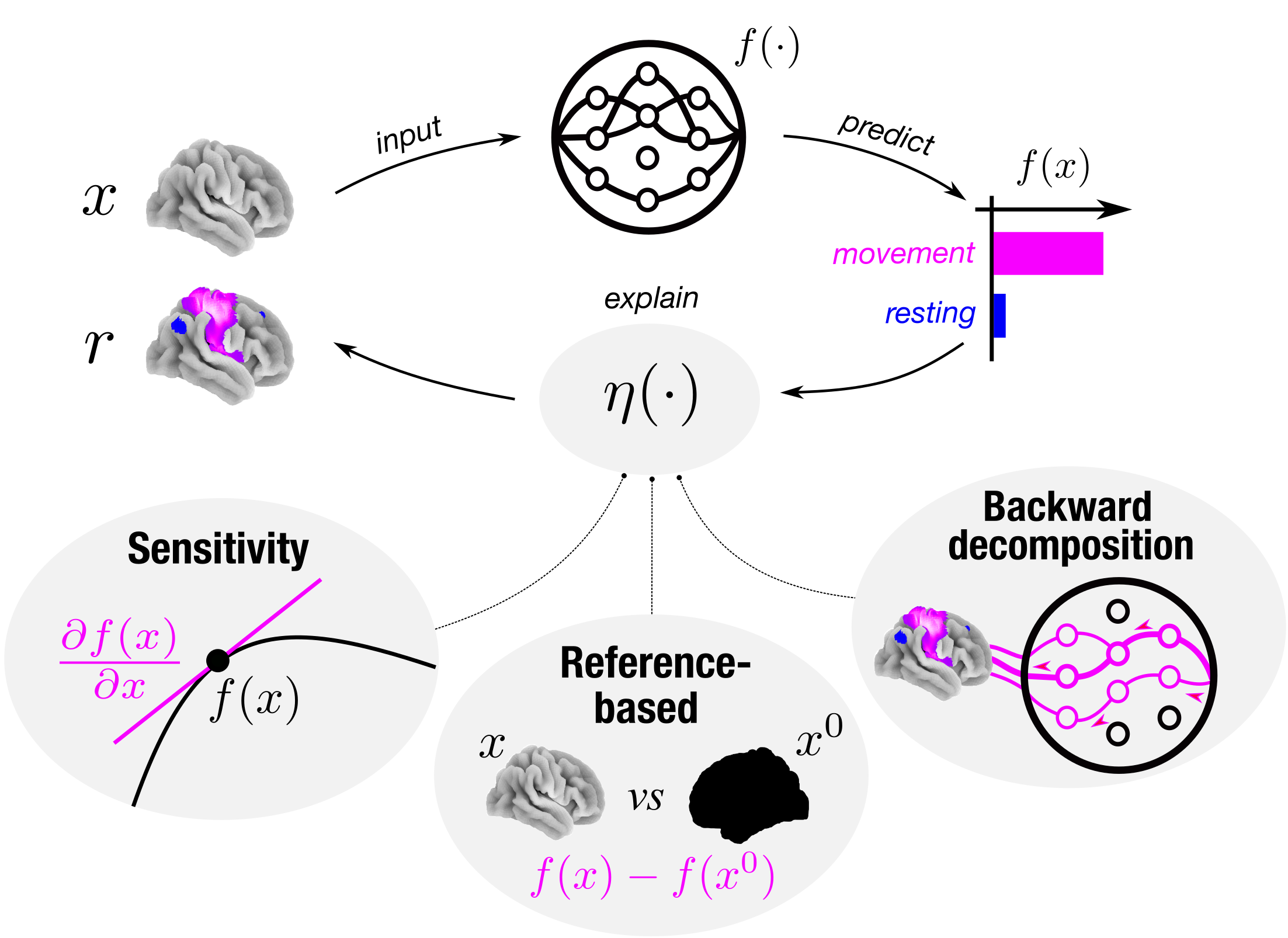}
\end{center}
\caption{Attribution approaches. All considered attribution approaches $\eta(\cdot)$ seek to explain a mental state decoding decision $f(x)$ of model $f(\cdot)$ by \textit{attributing} a relevance $r$ to each feature of the input $x$ for the decision. Backward decompositions attribute relevance to input features by decomposing the decoding decision in a backward pass through the model into the contributions of lower-level model units to the decision, up to the input space, where a contribution for each input feature can be defined. Reference-based attributions, by contrast, attribute relevance to input features by comparing the model's response to a given input to its response to a reference input $x^0$, often chosen to be neutral. Sensitivity analyses, on the other hand, attribute relevance to input features according to how sensitively a model's decoding decision responds to an input feature's value.}
\label{fig:methods:xai-approaches}
\end{figure}

We assume that the analyzed model represents some function $f(\cdot)$ mapping an
input $x \in \mathbb{R}^N$ to some output $f(x) \in \mathbb{R}$, such that: 
$f(\cdot): \mathbb{R}^N \xrightarrow{}\mathbb{R}$. In the following, we present 
a set of attribution methods $\eta(\cdot)$ that seek to provide insights into this mapping by 
attributing a relevance score $r_n$ to each input feature $n = 1, ..., N$, quantifying its 
contribution to $f(x)$, such that: $\eta(\cdot): \mathbb{R} \xrightarrow{}\mathbb{R}^N$. 

On a high level, the presented attribution methods can be divided into sensitivity
analyses, reference-based attribution methods, and backward decompositions (see Fig.\ \ref{fig:methods:xai-approaches}). 
Sensitivity analyses quantify $r$ by determining how sensitively $f(x)$ responds to
$x$. Reference-based attributions, by contrast, determine $r$ by contrasting the 
model’s response to a given input $x$ to its response to a reference input $b$. 
Backward decompositions, on the other hand, quantify $r$ by sequentially 
decomposing the model's output $f(x)$ in a backward pass through the model into 
the contributions of the lower-layer model units to the predictions, until the 
input space is reached and a contribution for each input feature can be defined.

\paragraph{Gradient Analysis~\citep{zurada_sensitivity_1994,simonyan_deep_2014}:} represents the most commonly used type of sensitivity analysis and defines $r_n$ as the partial derivative of 
$f(x)$ with respect to the input $x_n$, such that: $r_n = |\frac{\partial f(x)}{\partial x_n}|$. 
Relevance is thus assigned to those feature values to which $f(x)$ responds most sensitively.

\paragraph{SmoothGrad ~\citep{smilkov_smoothgrad_2017}:} represents 
an extension of Gradient Analysis to account for sharp fluctuations of the 
gradient $\nabla f(x) = \frac{\partial f(x)}{\partial x}$ at small scales,
which can otherwise lead to noise in the resulting attributions. 
SmoothGrad therefore first randomly draws $K$ 
samples (we set $K=50$) from the neighborhood of $x$ by adding random Gaussian
noise $\mathcal{N}(0,\sigma^2)$ with standard deviation $\sigma$ (we set $\sigma=1$) to $x$, and subsequently 
averages the resulting absolute partial derivatives for each random sample to obtain relevances
$r$: $r = \frac{1}{K} \sum_{1}^{K} |\nabla f(x + \mathcal{N}(0,\sigma^2))|$. 

\paragraph{InputXGradient~\citep{shrikumar_learning_2017}:}
represents another extension of Gradient Analysis, which multiplies the
gradient $\nabla f(x)$ by $x$, such that: $r = \nabla f(x) \times x$ (where $\times$ indicates the element-wise product). 
The intuition behind this approach comes from linear models, where the product of
input and model coefficient (here represented by the gradient) corresponds to the
total contribution of the associated feature to the model's output.

\paragraph{Guided Backpropagation~\citep{springenberg_striving_2015}:} represents an adaptation of Gradient Analysis tailored to CNN models that primarily use ReLU 
\cite{agarap_deep_2019} activation functions. It overrides gradients of ReLU activation functions in the computation of the gradient $\nabla f(x)$ such that only
non-negative gradients are backpropagated. 

\paragraph{Guided Gradient-weighted Class Activation Mapping (Guided GradCam)~\citep{selvaraju_grad-cam_2017}:} represents another type of  sensitivity analysis tailored to CNNs that combines Guided
Backpropagation with the GradCam~\citep{selvaraju_grad-cam_2017} technique. Specifically, 
GradCam first computes the gradient $\nabla f(x)$ with respect to the feature maps 
$A^k \in \mathbb{R}^D$ of the last convolutional layer $L$ closest to the model's output ($\frac{\partial f(x)}{\partial A^k}$) and then 
global-average pools the resulting gradients to obtain an importance weight $\alpha_k$ for each kernel $k \in L$: 
$\alpha_k = \frac{1}{D} \sum_{d=1}^D \frac{\partial f(x)}{\partial A_d^k} $. Conceptually, $a_k$ captures the importance of feature map $A^k$ for the decoded mental state.
Next, GradCam uses the importance weights $\alpha_k$ to combine the activation maps $A^k$ to an aggregate heatmap of relevances $r_A$: $r_A = \sigma (\sum_k \alpha_k A^k)$, 
where $\sigma$ represents the ReLU function ($\sigma(x) = max(0, x)$). 
Last, to obtain relevances $r$, Guided GradCam upsamples the relevances $r_A$ to the original input dimension and multiplies the upsampled maps with the relevance attributions of an application of Guided Backpropagation to $f(x)$ (see above).

\paragraph{Integrated Gradients~\citep{sundararajan_axiomatic_2017}:} 
 represents a reference-based attribution
method that assigns relevance $r$ by integrating the gradient $\nabla f(x)$ along a linear
trajectory in the input space, connecting the current input $x$ to some neutral reference
input $b$: $r^b = (x-b) \int_{\alpha=0}^{1}\frac{\delta f(b + \alpha(x-b))}{\delta
x}$. Integrated Gradients thus assigns relevance to input features according to how much
the model’s output changes when these features are scaled from the reference value to their
current value. In our analyses, we chose two reference inputs, namely, an all-zero input
$b^0$~\citep[as recommended in ][]{sundararajan_axiomatic_2017} as well as an average over
all inputs in the analyzed dataset $b^\mu$, and averaged over the two resulting attributions
to obtain relevance values $r$: $r = 0.5 r^0 + 0.5 r^\mu$.

\paragraph{Deep Learning Important FeaTures (DeepLift) DeepLift~\citep{shrikumar_learning_2017}:} represents another type of 
reference-based attribution method. Similar to Integrated 
Gradients, DeepLift determines relevances $r$ by comparing model
responses for a given input $x$ to the model's response to some neutral
reference input $b$. To this end, DeepLift defines a contribution score
$C_{\Delta x_n \Delta f(x)}$, describing the difference-from-reference response
$\Delta f(x) = f(x) - f(b)$ that is attributed to a difference-from-reference in the input 
$\Delta x = x - b$. Note that $\sum_{n=1}^N C_{\Delta x_n \Delta f(x)} = \Delta f(x)$. 
To compute these contribution scores, DeepLift uses  multipliers
$m_{\Delta x_n \Delta f(x)}$ that are defined as 
$m_{\Delta x \Delta f(x)} = \frac{C_{\Delta x \Delta f(x)}}{\Delta x}$ and thereby quantify the contribution of $\Delta x$ to $\Delta f(x)$, scaled by $\Delta x$. 
For any unit $x^{(l)}$ in model layer $l$ and any unit $x^{(l-1)}$ in the preceding layer
$l-1$, these multipliers can be computed as:
$m_{\Delta x^{(l-1)}_n \Delta f(x)} = \sum_j m_{\Delta x^{(l-1)}_n \Delta x^{(l)}_j} m_{\Delta x^{(l)}_j \Delta f(x)}$ (where $\Delta x^{(l)}_j$ indicates the difference in input feature $j$ of layer $l$ to its value for the reference input), 
in line with the chain rule. 
Relevance $r_n$ for input feature $n$ can then be obtained as: 
$r_n = C_{\Delta x_n \Delta f(x)} = \Delta x_n m_{\Delta x_n \Delta f(x)}$. 
In its basic formulation, DeepLift uses two rules to compute contribution scores: 
The linear rule applies to dense and convolution layers, which compute 
$a = w^0 + \sum_{n} w_n x_n$ (where $w^0$ and $w$ indicate bias and weights and 
$x$ the input), and defines $C_{\Delta x_n \Delta a} = \Delta x_n w_n $ and
accordingly $m_{\Delta x_n \Delta a} = \frac{C_{\Delta x_n \Delta a}}{\Delta x_n}$. 
The rescale rule applies to all non-linear transformations $\sigma(a)$ 
(e.g., ReLU or sigmoid functions) and defines $C_{\Delta x \Delta \sigma(a)} = \Delta \sigma(a)$ 
and accordingly $m_{\Delta x \Delta \sigma(a)} = \frac{C_{\Delta x \Delta \sigma(a)}}{\Delta x}$.

\paragraph{DeepLift SHapley Additive exPlanation (DeepLift SHAP)~\citep{lundberg_unified_2017}:} combines DeepLift with SHAP, a method for
computing Shapley values~\citep{shapley_value_1952} for a conditional expectation
function of the analyzed model. Specifically, SHAP values attribute to each input feature 
the change in expected model prediction conditioned on a feature of interest. To 
approximate SHAP values using DeepLift for a given input $x$, DeepLift SHAP draws $K$ (we set $K=50$) 
random samples from the data (to approximate the set of other possible feature
coalitions) and averages over the DeepLift attributions for each random sample, 
when treating the random sample as a reference input $b$.

\paragraph{Layer-wise relevance propagation (LRP)~\citep{bach_pixel-wise_2015}:} represents a backward decomposition method. Let $i$ and $j$ be the indices of two models units in successive model layers $l$ and $l+1$ and $r_j^{(l+1)}$ the relevance of unit $j$ for $f(x)$. To
redistribute relevance between layers, several rules have been proposed
\citep{bach_pixel-wise_2015,montavon_layer-wise_2019,kohlbrenner_towards_2020}, 
which generally follow from $r_i^{(l)} = \sum_j \frac{a_i w_{ij}}{\sum_i a_i
w_{ij}} r_j^{(l+1)}$ (where $a$ and $w$ represent the input and weight of model unit 
$i$ in layer $l$). Importantly, LRP conserves relevance between layers, such that: $\sum_n r_n = \sum_i r_i^{(l)} = \sum_j r_j^{(l+1)} = f(x)$. 
In line with the recommendations by~\citep{montavon_layer-wise_2019}, 
we use a composite of relevance redistribution rules, namely, the LRP-0 rule 
($r_i^{(l)} = \sum_j \frac{a_i w_{ij}}{\sum_i a_i  w_{ij}} r_j^{(l+1)}$) 
for the dense output layer and the LRP-$\gamma$ rule 
($r_i^{(l)} = \sum_j \frac{a_i (w_{ij} + \gamma w_{ij}^+)}{\sum_i a_i (w_{ij} + \gamma  w_{ij}^+)} r_j^{(l+1)}$, 
where $\gamma$ controls the positive contributions to $r$ (we set $\gamma=0.25$)) for all convolution layers. 

%% file: sections/methods/statistical-modelling.tex
\subsection{Statistical modelling}
\label{methods_statistical_modelling}

For the statistical comparison of attribution methods on any metric, we estimate linear regression models that include one binary indicator variable for all attribution methods other than DeepLift, which we treat as unmodelled baseline. We fit all regression models in a Bayesian framework by the use of the Bayesian Model-Building Interface~\citep[bambi 0.9.0;][]{capretto_bambi_2022} and by sampling four chains per parameter ($5,000$ samples per chain after $5,000$ discarded tuning samples) using the Markov chain Monte Carlo No-U-Turn-Sampler~\citep[NUTS; ][]{hoffman_no-u-turn_2014} with bambi's automatically generated priors. We determine a method as meaningfully different from DeepLift on the analyzed metric if the estimated $94\%$ highest-density interval of the method's coefficient does not include $0$. All posterior traces are checked for convergence according to the Gelman–Rubin statistic ($|\hat{R} - 1| <.01$).

%% file: sections/methods/code-data-availability.tex
\subsection{Code and data availability}

All custom code that we use for the analyses of this  study as well as the trial-level BOLD maps are available at:
\href{https://github.com/athms/interpreting-brain-decoding-models}{github.com/athms/interpreting-brain-decoding-models}.

%% file: sections/results/hyperopt.tex
The target of our decoding analyses was to correctly decode the mental state
associated with each trial-level BOLD map (namely, "heat" and "rejection"
for the heat-rejection dataset, "left foot (lf)", "right foot (rf)", 
"left hand (lh)", "right hand (rh)", and "tongue (t)" for the MOTOR dataset,
and "body", "faces", "places", and "tools" for the WM dataset; see section \ref{methods_experiment_tasks}).

\subsection{Hyper-parameter optimization}
\label{results_hyperopt}

To determine a set of model and optimization hyper-parameters for each dataset, we performed a grid search over 144 unique parameter sets (see section \ref{methods_hyperopt}) and evaluated the performance of each of these configurations in a 
three-fold cross-validation procedure, using each dataset's training data (see section \ref{methods_data_split}).
We then determined the best-performing configuration by computing 
each configuration's mean decoding error $\epsilon$ (defined as $1$ minus the fraction of
correctly decoded trial-level BOLD maps) in the
training and validation datasets over the three folds ($\epsilon^T$ 
and $\epsilon^V$ respectively) and assigning a performance score
$\lambda_i$ to each configuration $i$: 
$\lambda_i = \epsilon^{V}_i + |\epsilon^{V}_i - \epsilon^{T}_i|$. 
This score quantifies how well a model performed in the validation data 
when also accounting for the difference in model performance to the training data.
Accordingly, we defined the best performing configuration for each dataset as the
one with the lowest $\lambda_i$ (see Table 1).

\begin{table}[]
\caption{Best-performing model and optimization configurations. For each dataset, the number of 3D-convolution layers, number of kernels per convolution layer, size of the kernels, training mini-batch size (BS), learning rate (LR), and dropout rate are shown for the configuration with lowest $\lambda_i$ (indicating best performance; see section \ref{results_hyperopt})}
\centering
\begin{adjustbox}{width=0.8\textwidth}
\begin{tabular}{lrrrrrr}
\multicolumn{7}{c}{\multirow{2}{*}{}}                                                               \\ \hline
\textbf{Dataset}                         & \textbf{\# Layers}                         & \textbf{\# Kernels}                         & \textbf{Kernel size}                         & \textbf{BS}                        & \textbf{LR}                        & \textbf{Dropout}                        \\ \hline
heat-rejection                           & 4                                          & 8                                           & 3                                            & 32                                 & 0.001                              & 0.5                                     \\
MOTOR                                    & 3                                          & 8                                           & 3                                            & 32                                 & 0.001                              & 0.5                                     \\
WM                                       & 4                                          & 16                                          & 5                                            & 64                                 & 0.001                              & 0.5                                     \\ \hline
\end{tabular}
\end{adjustbox}
\label{table:best-performing-models}
\end{table}

%% file: sections/results/decoding-performance.tex
\subsection{Models accurately decode mental states}
\label{results_decoding-performance}

\begin{figure}[!t]
\begin{center}
\includegraphics[width=0.9\linewidth]{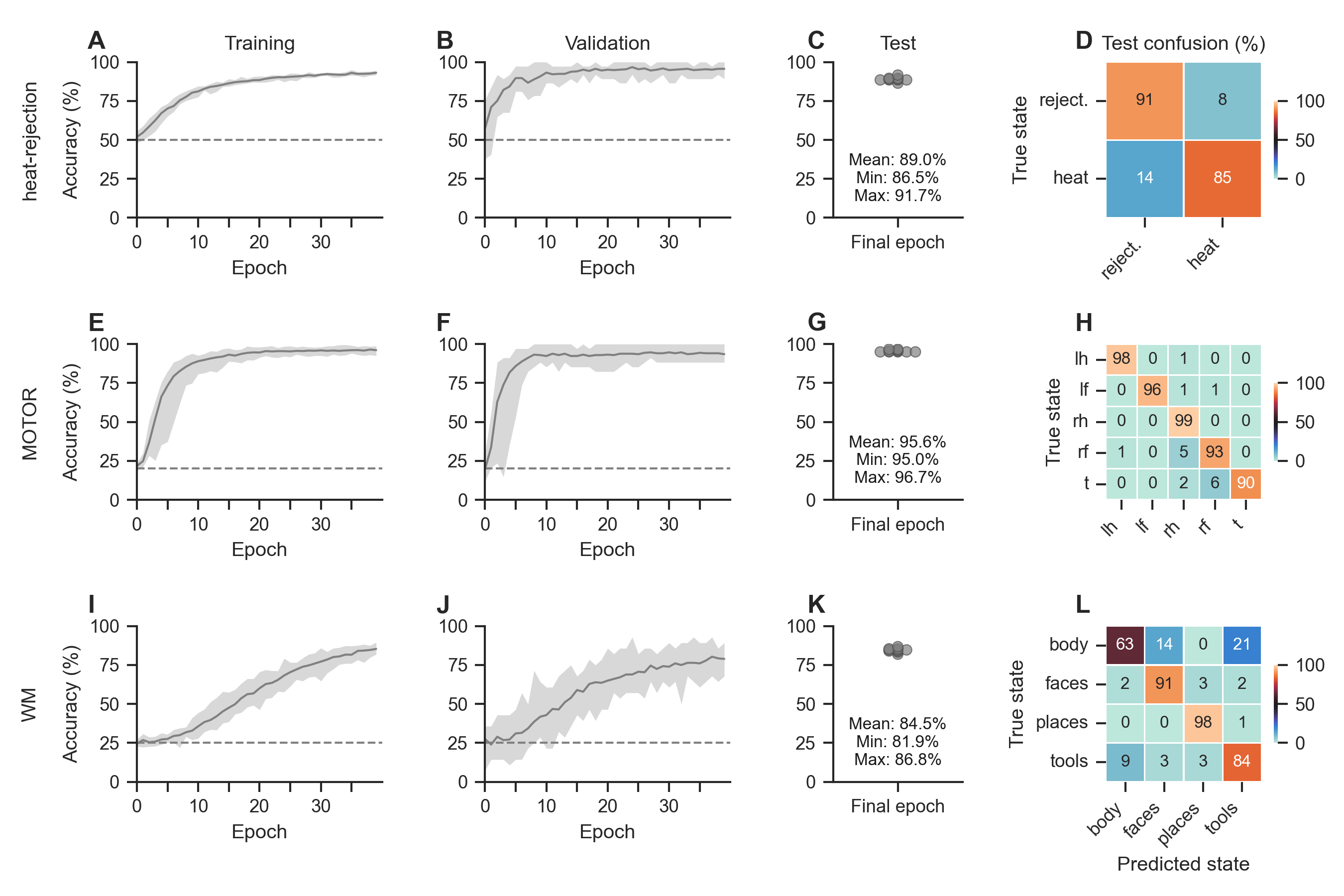}
\end{center}
\caption{Decoding performance. For each dataset, we trained ten identical variants 
of each datasets' model and optimization configuration (see section \ref{results_decoding-performance} 
of the main text for details on hyper-parameter selection) and solely varied the 
random seed between training runs and the random split of the data into training and validation 
datasets (95\% and 5\% of the full training data respectivelyy). A-B,E-F,I-J: The configurations performed well 
in decoding the mental states from the trial-level maps in the training (A,E,I) and 
validation datasets (B,F,J). Lines indicate mean decoding accuracies with 
shaded areas indicating respective minimum and maximum decoding accuracies across 
the ten training runs. Dashed line indicates chance accuracy. C-D,G-H,K-L: The final models also performed well in decoding 
the mental states of the left-out test datasets (C,G,K) with overall low average confusion rates (D,H,L). Scatter points indicate the final decoding accuracies in the test data.}
\label{fig:results:decoding-performance}
\end{figure}

Several recent findings in DL research have demonstrated that DL model performances are strongly dependent on many non-deterministic aspects of their training, 
such as, random weight initializations and random shufflings of the data during training~\citep{thomas_challenges_2021,lucic_are_2018,henderson_deep_2018}. To this end, we performed ten training runs with the best-performing model and optimization configuration of each dataset (see section \ref{results_hyperopt}), with different random seeds and  
training/validation data splits per run.
For each run, we divided the dataset's training data (see section \ref{methods_data_split}) into new training and
validation datasets by randomly selecting $5\%$ of the trial-level BOLD maps as
validation data and using the remaining maps for training. We then trained models for $40$ epochs on the training data (Fig.\ \ref{fig:results:decoding-performance} A-B,E-F,I-J) before evaluating the model's decoding performance in the left-out test dataset (containing the data of 
every 5th subject of the full dataset; see Fig.\ \ref{fig:results:decoding-performance} C-D,G-H,K-L). 

The models performed well in decoding the mental states of each dataset, with average test decoding accuracies of 89.0\% [86.5\%, 91.7\%] (heat-rejection), 
95.6\% [95\%, 96.7\%] (MOTOR), and 84.5\% [81.9\%, 86.8\%] (WM) (reported as mean [min, max]) (Fig.\ \ref{fig:results:decoding-performance} C,G,K). 
We also computed average test confusion rates over the ten training runs and found that the models exhibited little confusion between the mental states (Fig.\ \ref{fig:results:decoding-performance} D,H,L)).

%% file: sections/results/brain-maps.tex
\subsection{Explanations of sensitivity analyses are biologically more plausible}
\label{results_brain-maps}

\begin{figure}[!t]
\begin{center}
\includegraphics[width=0.9\linewidth]{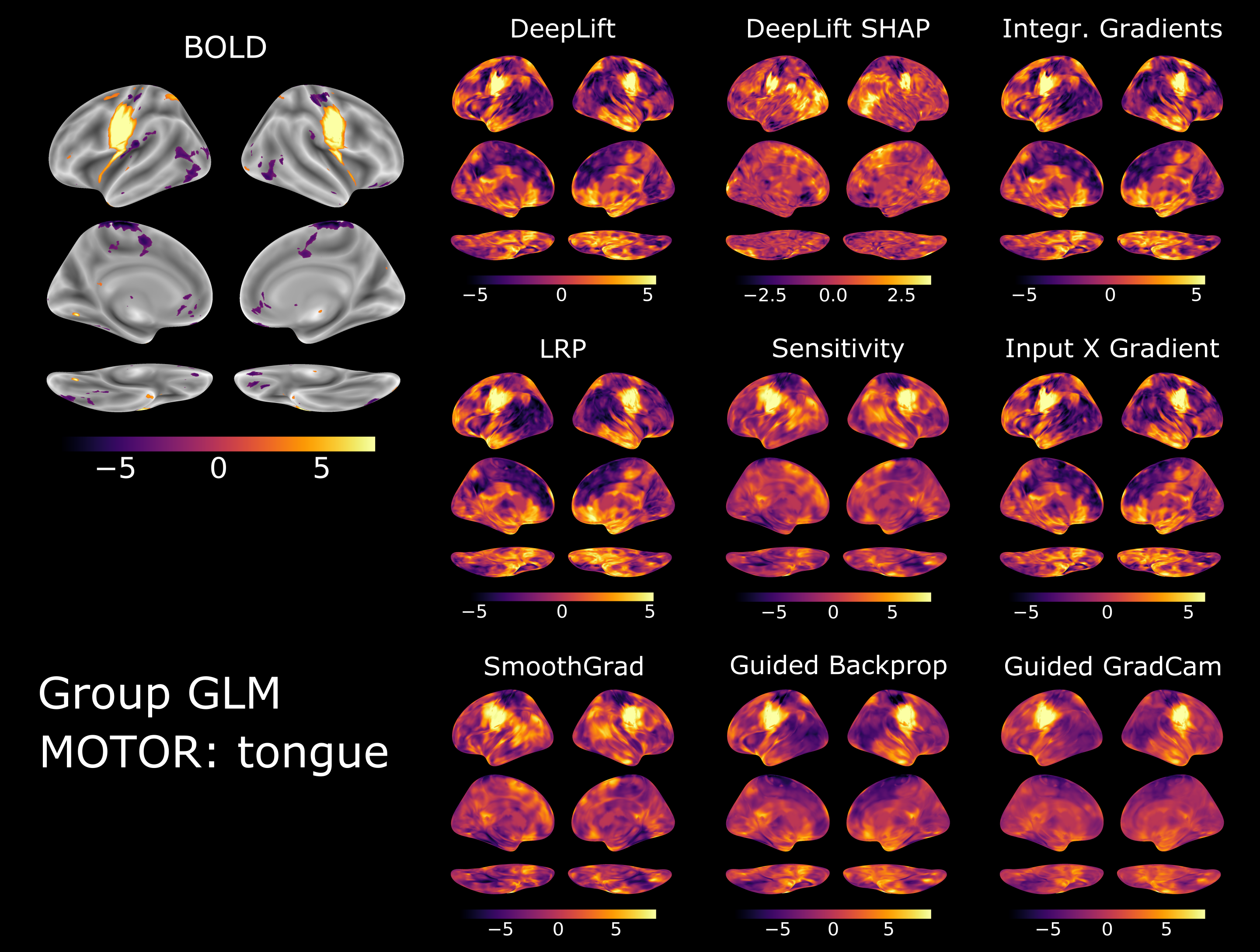}
\end{center}
\caption{Group-level brain maps for "tongue" movements of the MOTOR dataset. 
For each dataset, we performed a two-stage random effects GLM analysis of the trial-level 
BOLD maps as well as the attributions of each attribution method for test trial-level BOLD maps. 
Brain maps show the resulting Z-scores of a contrast between the "tongue" state and all other mental states of 
the MOTOR dataset. All brain maps are projected onto the inflated cortical surface of the FsAverage template~\citep{fischl_freesurfer_2012}. We thresholded 
the group-level BOLD contrast map at a false positive rate of $0.01$.}
\label{fig:results:brain-maps}
\end{figure}

As the trained models performed well in decoding the mental states of 
the three datasets (see Fig.\ \ref{fig:results:decoding-performance}), we proceeded to compare the attribution methods' effectiveness in providing insight into the models' learned mappings between brain activity and mental states. 
To this end, we interpreted the decoding decisions of each of the ten trained model instances per dataset (see section \ref{results_decoding-performance}) with each attribution method (see section \ref{methods_xai_approaches}) for the trial-level BOLD maps of the corresponding test dataset (see section \ref{methods_data_split}). 
Importantly, we always interpreted the models' decoding decision 
for the actual mental state associated with each trial-level BOLD map.
This analysis resulted in a dataset of ten attribution maps (one per model instance) for each attribution method and trial-level BOLD map. 

To aggregate these attribution data over the ten model training runs, we performed a standard two-stage GLM analysis by first computing a subject-level GLM analysis of the trial-level attribution maps and subsequently aggregating the subject-level attribution maps in a random-effects group-level analysis (for details on the GLM analysis, see section \ref{methods_subject_group_glm}). 
Importantly, the subject-level analysis included additional binary nuissance variables for the ten model training runs as well as the sum of the attribution values of each trial-level attribution map (as attribution sums can vary between decoding decisions, e.g., due to varying model decoding certainty). 

To also identify the set of voxels whose activity we would expect to be associated with each mental state 
in a standard analysis of the BOLD data, we repeated this GLM analysis procedure for the trial-level BOLD maps of each
dataset (without the additional nuisance regressors).

Figure \ref{fig:results:brain-maps} provides an overview of the resulting group-level BOLD and attribution maps for the "tongue" movement state of the MOTOR dataset. 
For this state, the attribution methods correctly attributed high relevance to those voxels in the ventral premotor and primary motor cortex that also showed high activity in the group-level analysis of the BOLD data.

\begin{figure}[!t]
\begin{center}
\includegraphics[width=0.9\linewidth]{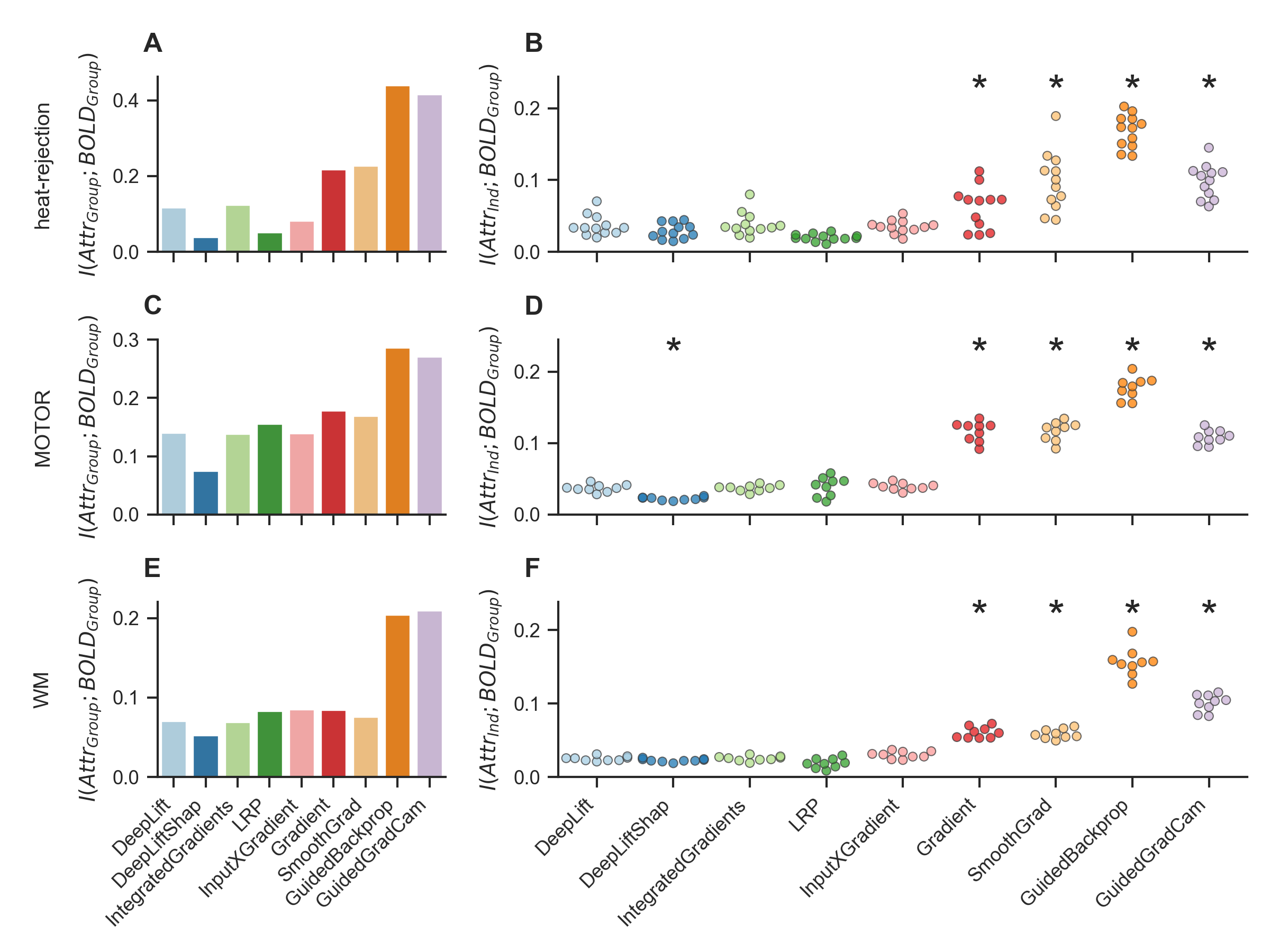}
\end{center}
\caption{Similarity of attribution maps to group-level BOLD maps. To estimate similarity, 
we compute the mutual information between the group-level BOLD maps and the group- (A,C,E) and subject-level (B,D,F) 
attribution maps for each mental state. 
A,C,E: The group-level maps of Guided Backpropagation and Guided GradCam exhibit overall 
higher mutual information (i.e., similarity) with the group-level BOLD maps than the 
group-level maps of the other attribution methods. Bar heights indicate mean mutual 
information over the mental states of a dataset. B,D,F: Similarly, the subject-level 
maps of Guided Backpropagation, Guided GradCam as well as Gradient Analysis and SmoothGrad 
exhibit higher mutual information with the group-level BOLD maps than the subject-level 
maps of the other attribution methods. Scatter points indicate mean mutual information per subject. Black stars indicate that the distribution of subject means is meaningfully different from the distribution of the DeepLift method (for analysis details, see section \ref{methods_statistical_modelling}). Colors indicate interpretation methods.}
\label{fig:results:mutual-info}
\end{figure}

As can be seen, it is generally difficult to discern the quality of the various attributions 
by visual inspection alone. For this reason, we next took a quantitative approach to 
analyzing how well the attributions align with the 
results of the GLM analysis of the BOLD data by computing the average mutual information~\citep{kraskov_estimating_2004} between the group-level attribution maps and the 
corresponding group-level BOLD maps for the same mental state 
(see Fig.\ \ref{fig:results:mutual-info} A,C,E). Note that we chose mutual information 
as a similarity measure because the association between the group-level BOLD and 
attribution maps does not need to be linear. For example, it is possible that a DL 
model learns to identify a mental state through the activity of voxels that are 
meaningfully more active in this state as well as the activity of voxels that are 
meaningfully less active, resulting in an attribution map that assigns high relevance 
to voxels that exhibit high positive and negative values in a GLM analysis of the BOLD data~\citep{thomas_evaluating_2021}.

This analysis revealed that the group-level attribution maps of Guided Backpropagation~\citep{springenberg_striving_2015} 
and Guided GradCam~\citep{selvaraju_grad-cam_2017}, two types of sensitivity
analysis (see section \ref{methods_xai_approaches}), are more similar to the group-level BOLD maps than those of the other attribution methods (as indicated by average higher mutual information scores), while the group-level maps of Gradient Analysis~\citep{simonyan_deep_2014} 
and SmoothGrad~\citep{smilkov_smoothgrad_2017}, again two types of sensitivity analyses, exhibit less, but still comparably high similarity, to the group-level BOLD maps when compared to the remaining attribution methods
(Fig.\ \ref{fig:results:mutual-info} A,C,E).

We also tested how well the subject-level attribution maps of each attribution method align with the group-level BOLD maps, as the trained models can draw from their knowledge about the group of subjects in their training data when decoding the trial-level BOLD maps of the test datasets. 
To this end, we computed the mutual information between the subject-level attribution maps and the corresponding group-level BOLD map of the same mental state and regressed the average mutual information per subject onto a set of binary dummy variables indicating the attribution methods (for details on the regression analysis, see section \ref{methods_statistical_modelling}).
This analysis showed that the subject-level attribution maps of Guided Backpropagation, Guided GradCam, SmoothGrad, and Gradient Analysis, all variants of sensitivity analysis, are generally more similar to the group-level BOLD maps than the subject-level attribution maps of the other attribution methods (Fig.\ \ref{fig:results:mutual-info} B,D,F). 

%% file: sections/results/faithfulness.tex
\subsection{Explanations of reference-based attributions and backward decompositions are more faithful}
\label{results_faithfulness}

\begin{figure}[!t]
\begin{center}
\includegraphics[width=0.9\linewidth]{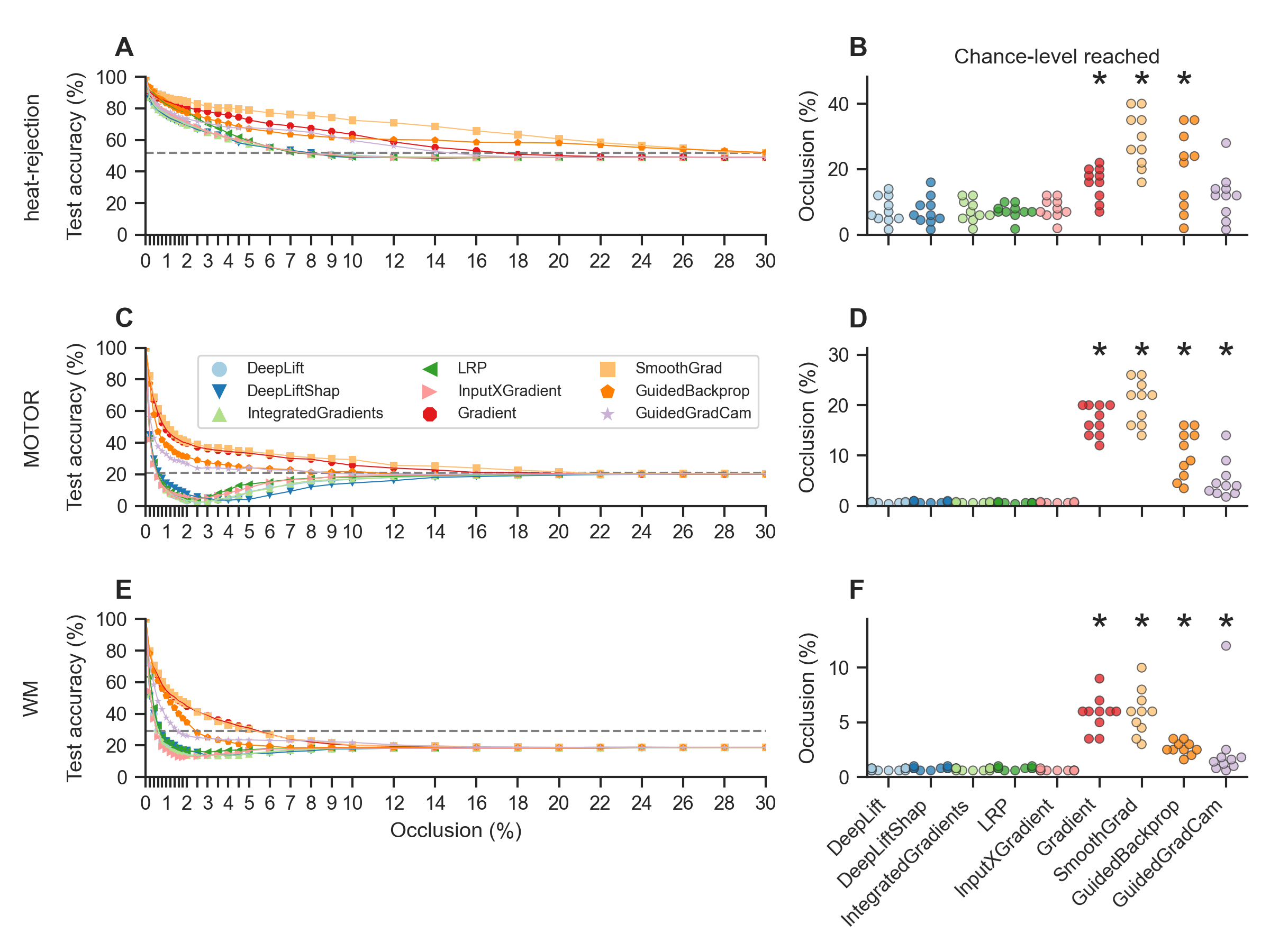}
\end{center}
\caption{Attribution faithfulness. We estimated explanation faithfulness of an attribution method by 
repeatedly evaluating the models' test decoding accuracy when occluding 
different fractions of the input voxels based on the relevance assigned to 
them by the attribution method, such that $0\%$ occlusion indicates that 
no voxel values were occluded while $50\%$ indicates that the values of all 
input voxels were occluded that received the
highest $30\%$ of relevance values. For each 
attribution method and trained model, we recorded the occlusion rate at which the models' test decoding accuracy first dropped to chance-level, 
indicating that all information has been removed from the data that the model uses to correctly identify the mental states. A,C,E: DeepLift, 
DeepLift SHAP, Integrated Gradients, LRP, and InputXGradient  
exhibit higher attribution faithfulness than Gradient Analysis, SmoothGrad, 
Guided Backpropagation, and Guided GradCam, as the models' test decoding 
accuracies decrease more rapidly with increasing occlusion rates for these methods. Lines indicate mean test decoding accuracies 
over the ten training runs of each model and optimization configuration. B,D,F: Accordingly, 
model test decoding accuracies also drop to chance-level at lower occlusion 
rates for these approaches. Scatter points represent model training runs and black stars indicate that the distribution of occlusion rates is meaningfully different from the distribution of the DeepLift method (for analysis details, see section \ref{methods_statistical_modelling}). Colors 
indicate interpretation methods.}
\label{fig:results:faithfulness}
\end{figure} 

In addition to understanding how the explanations of each attribution method compare to the results of a standard GLM analysis of the BOLD data, 
we were interested in understanding how well they  perform at capturing 
the decision process of the trained models. 
To this end, we analyzed their explanation faithfulness~\citep{samek_explaining_2021,samek_evaluating_2017}.
An explanation can generally be considered as being faithful if it correctly identifies those features of the input that are most relevant for the model's decoding decision. 
Accordingly, we tested whether removing those voxels from the input that received high relevance by an attribution method (by setting their values to 0) affects the model's ability to correctly identify the mental states. 
To quantify faithfulness, we repeated this analysis for different occlusion rates, from $0\%$ (indicating no occlusion) to $50\%$
(indicating that those voxels are occluded that received the 
highest $50\%$ of relevance values) (Fig.\ \ref{fig:results:faithfulness} A,C,E) and recorded the occlusion 
rate at which the model's decoding accuracies first dropped to chance level, 
indicating that all information has been removed from the data that the model 
requires to accurately identify the mental states 
(Fig.\ \ref{fig:results:faithfulness} B,D,F). If an attribution method has high explanation faithfulness, the model's decoding accuracy should drop to chance level at lower occlusion rates when compared to methods with lower faithfulness.

Overall, this analysis revealed that reference-based attributions and backward decompositions, namely, DeepLift, DeepLift SHAP,
Integrated Gradients, LRP generally exhibit higher explanation faithfulness than the tested sensitivity analyses because the models' decoding 
decisions dropped to chance level at lower occlusion rates for these methods than for the others (with the exception of the InputXGradient~\citep{shrikumar_learning_2017}
method whose explanations exhibited similar faithfulness to the tested backward decompositions and reference-based attributions; 
Fig.\ \ref{fig:results:faithfulness}).

%% file: sections/results/sanity-checks.tex
\subsection{Sanity checks for attribution methods}
\label{results_sanity-checks}

\begin{figure}[!t]
\begin{center}
\includegraphics[width=0.9\linewidth]{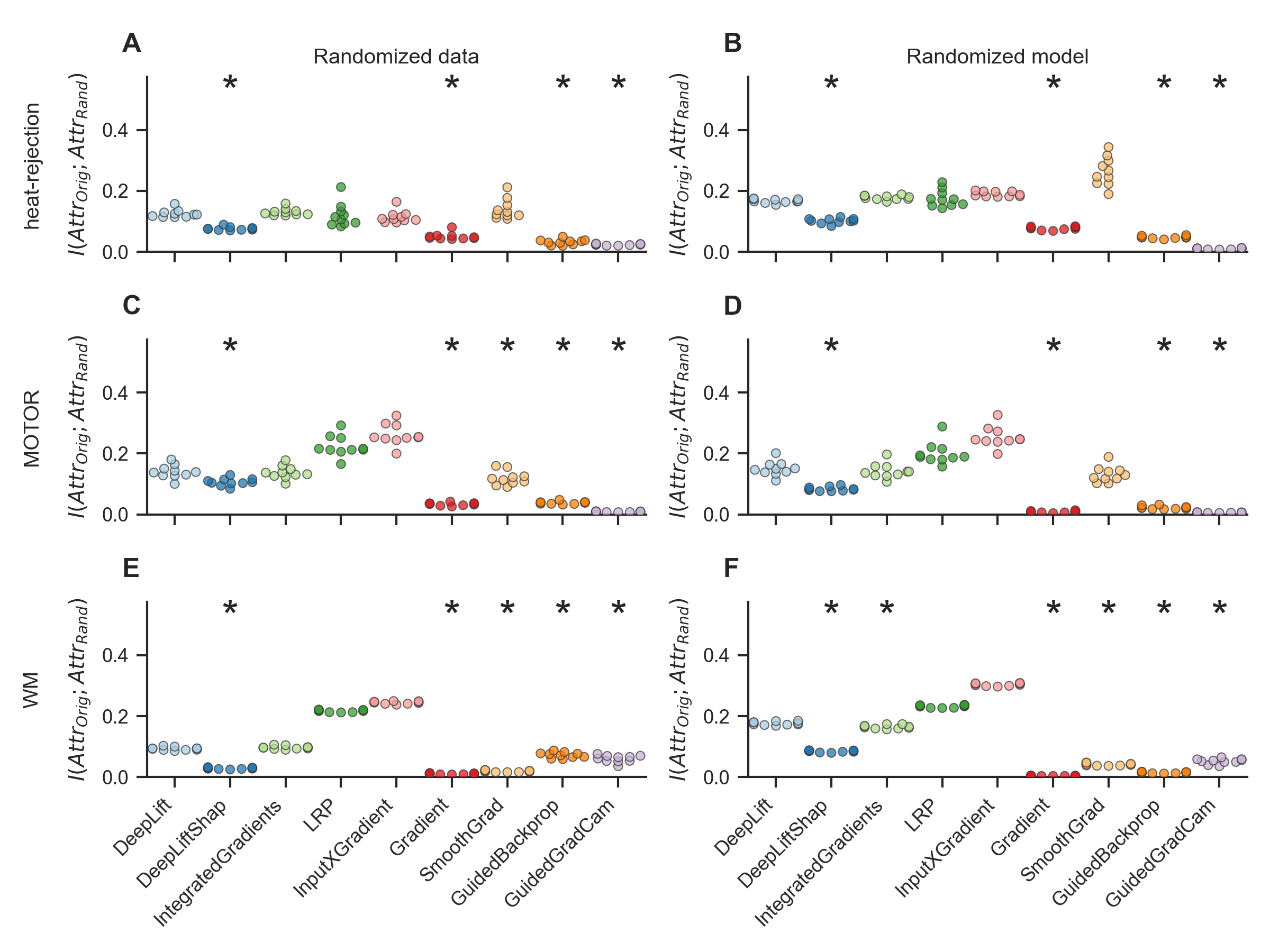}
\end{center}
\caption{Sanity checks for attribution methods. We performed two sanity for each attribution method by comparing a method's original attributions to its attributions for the same data when interpreting the mental state decoding decisions of a model variant trained on a version of the training data with randomized labels (data randomization test; A, C, E) and model with randomized weights (model randomization test; B, D, F). If the attributions of an attribution method 
are sensitive to the characteristics of the input data and model, its attributions for the original model should be different from its attributions for the model trained on 
randomized data as well as the model with random weights (leading to low mutual information scores). Overall, DeepLift SHAP, Gradient Analysis, Guided Backpropagation, and Guided GradCam performed better than the other methods in both tests (as indicated by generally lower mutual information scores), while DeepLift and Integrated Gradients also performed comparably well. 
Scatter points indicate mean mutual information for each model training run. Black stars indicate that the distribution of mean mutual information scores is meaningfully lower than the distribution of the DeepLift method (for analysis details, see section \ref{methods_statistical_modelling}). Colors 
indicate interpretation methods.}
\label{fig:results:sanity-checks}
\end{figure}

Last, we performed two sanity checks for attribution methods, as recently 
proposed by~\citet{adebayo_sanity_2018}, to test the overall scope and quality of their explanations. Specifically,~\citet{adebayo_sanity_2018} propose two simple 
tests to test whether the explanations of an attribution method
are specific to the tested model and data by testing how much the explanations 
change when the method is applied to a model with the same architecture but random weights 
(the model randomization test) or a model trained on a version of 
the data with randomly permuted labels (the data randomization test). 
If the explanations are dependent on the specific 
parameters of the model, they should differ between the originally trained models and models with random weights. 
If the explanations are similar, however, the attribution method can be considered 
insensitive to the studied model and therefore not well-suited to capture the 
models' decision process. 
Similarly, if the explanations of an attribution 
method account for the labeling of the data, it should produce explanations 
that are different between models model trained on the original dataset and models trained on a version of the data with randomly shuffled labels. 
If the explanations are similar, however, the attribution method can be considered independent 
of the labeling of the data and therefore not well-suited to understand the 
model's learned mapping between these labels and the input data.

In line with the data randomization test, we first trained each model configuration on a version of its original training dataset 
with randomly shuffled mental state labels. The models were able to correctly decode the randomly 
shuffled mental state labels in their training data after $2,000$ training epochs, 
achieving decoding accuracies of $85.1\%$, $98.7\%$, and $98.4\%$ for the heat-rejection, 
MOTOR, and WM datasets respectively (Appendix Fig.\ \ref{appendix:fig:results:randomized-labels}). Importantly, the models' validation decoding accuracies were still close to chance (namely, $45.9\%$, $23.5\%$, and $25\%$ for the heat-rejection, MOTOR, and WM datasets 
respectively; Appendix Fig.\ \ref{appendix:fig:results:randomized-labels}), indicating that 
the models memorized the random associations between labels and training data. 
When comparing the attributions of each attribution method for the test mental state decoding decisions of the models trained on the randomized and original training data, we found that the attributions of DeepLift SHAP, Gradient Analysis, SmoothGrad, Guided Backpropagation, and Guided GradCam were generally more dissimilar between the two models than those of the other attribution methods, indicating stronger dependence of their attributions on the characteristics of the data.

Similarly, in line with the model parameter randomization test, we also interpreted 
the test mental state decoding decisions of a randomly initialized variant of each 
datasets' model configuration. 
As for the data randomization test, we compared the resulting attributions of each attribution method to the attributions for the originally trained models. Again, DeepLift SHAP, Gradient analysis, Guided Backpropagation, and Guided GradCam produced attributions 
that showed stronger dependence on the characteristics of the analyzed models, when compared to the other methods, as their attributions for the randomized and originally trained models were more dissimilar.

%% file: sections/discussion.tex
With this work, we provide insights into the explanation performance of prominent types of attribution methods, namely, sensitivity analyses, backward decompositions, and reference-based attributions (see section \ref{methods_xai_approaches}), in mental state decoding analyses with DL models. To evaluate explanation performances, we use a diverse set of criteria: First, we evaluate the biological plausibility of the explanations by comparing them to the results of a standard GLM analysis of BOLD data, which seeks to identify all voxels whose activity pattern is associated with the mental states. We find that sensitivity analyses, such Guided Backpropagation, Guided GradCam, Gradient Analysis, and SmoothGrad, provide explanations that are more similar to the results of the GLM analysis, and thereby biologically more plausible, than the explanations of the tested backward decompositions and reference-based attributions. Second, we evaluate whether the explanations accurately capture the models' decision process by testing whether they accurately identify all voxels of the input that the models rely on to identify the mental states. We find that backward decompositions and reference-based attributions, such as DeepLift, DeepLift SHAP, Integrated Gradients, and LRP, provide explanations that are more faithful than those of the tested sensitivity analyses. Last, to test whether the methods' explanations are in fact sensitive to the analyzed model and data, we perform two sanity checks for attribution methods~\citep[as suggested by][]{adebayo_sanity_2018} and find that  Gradient Analysis, Guided Backpropagation, Guided GradCam, DeepLift SHAP, DeepLift, and Integrated Gradients perform consistently well in both sanity checks, providing explanations that are sensitive to the characteristics of the analysed model and data.

\subsection{Biological plausibility vs. explanation faithfulness}

Our findings demonstrate a gradient between two key characteristics for the interpretation of mental state decoding decisions: attribution methods that provide highly faithful explanations, by capturing the model's decision process well, also provide explanations that are biologically less plausible, because they do not necessarily identify all voxels whose activity patterns are associated with the mental states, when compared to interpretation methods with less explanation faithfulness. To make sense of this finding, it is important to remember that functional neuroimaging data generally exhibit strong spatial correlations, such that individual mental states are often associated with the activity of large clusters of voxels. DL models trained to identify these mental states from neuroimaging data will likely view some of this activity as redundant, as the activity of a subset of those voxels suffices to correctly identify the mental states. In these situations, any explanation of an attribution method with perfect faithfulness will not identify all voxels of the input whose activity is in fact associated with the decoded mental state, but solely the subset of voxels whose activity the model used as evidence for its decoding decision. Accordingly, attribution methods with high explanation faithfulness, such as backward decompositions and reference-based attributions, do not necessarily produce explanations that align well with the results of a standard GLM analysis of the BOLD data. By contrast, we found that sensitivity analyses, such as Guided GradCam, Guided Backpropagation, Gradient Analysis, and SmoothGrad, produce explanations that are less faithful but more in line with the results of a standard GLM analysis. Sensitivity analyses are less concerned with identifying the specific contribution of each input voxel to a decoding decision and instead focus on identifying how sensitively a model's decision responds to (i.e., changes with) the activity of each voxel. With this perspective, sensitivity analyses identify a broader set of voxels whose activity the model takes into account when decoding the mental state, resulting in explanations that seem biologically more plausible because the set of identified voxels is more similar to that of standard analysis approaches for BOLD data, which seek to identify voxels whose activity pattern is associated with the mental state.

\subsection{Recommendations for interpretation methods in mental state decoding}

Based on these findings, we make a twofold recommendation for the application of attribution methods in mental state decoding: 

If the goal of a mental state decoding analysis is to understand the decision process of the decoding model by identifying the parts of the input that are most relevant for the model's decision, we generally recommend the application of backward decompositions or reference-based attributions. In particular, we recommend DeepLift, DeepLift SHAP, and Integrated Gradients because their explanations have shown overall high faithfulness in our analyses, while also performing well in the two sanity checks. 

By contrast, if the goal of a mental state decoding analysis is to understand the association between the BOLD data and studied mental states, we recommend the application of sensitivity analyses, as these have shown to produce explanations with comparably high similarity to the results of a standard GLM analysis of the data when compared to reference-based attributions and backward decompositions. Particularly, for CNN models with ReLU activation functions, we recommend Guided Backpropagation and Guided GradCam as their explanations exhibit the overall highest similarity to the results of a standard GLM analysis of the BOLD data in our analyses, while also performing well in the two sanity checks. For models without ReLU activation functions, we recommend Gradient Analysis and SmoothGrad, as their explanations also have comparably high similarity to the results of the GLM analysis (especially on the level of individual subjects), while also performing well in the two sanity tchecks. 

\subsection{Caution in the interpretation of complex models}

Last, we would like to advocate for caution in any interpretation of the mental state decoding decisions of DL models. DL models have an unmatched ability to learn from and represent complex data. Accordingly, their learned mappings between input data and decoding decisions can be highly complex and counterintuitive. For example, recent empirical work has shown that DL methods trained in mental state decoding analyses can identify individual mental states through voxels that exhibit meaningfully stronger activity in these states as well as voxels with meaningfully reduced activity in these states, leading to explanations that assign high relevance scores to voxels that receive both positive and negative weights in a standard GLM contrast analysis of the same BOLD data~\citep{thomas_evaluating_2021}. To understand how a model's weighting of the input in its decoding decision relates to the characteristics of the input data, it is therefore essential to compare the explanations of any attribution method to the results of standard analyses of the BOLD data~\cite[e.g., with linear models;][]{friston_statistical_1994,kriegeskorte_information-based_2006,grosenick_interpretable_2013} as well as related empirical findings~\cite[e.g., as provided by NeuroSynth;][]{yarkoni_large-scale_2011}.

Similarly, a wealth of recent empirical work in machine learning research has demonstrated that DL models are prone to learning simple shortcuts (or confounds) from their training data, which do not generalize to other datasets~\cite[for a detailed discussion, see][]{geirhos_shortcut_2020}. A prominent example is a study that trained DL models to identify pneumonia from chest X-rays \citep{zech_variable_2018}. While the models performed well in the training data, comprising X-rays from few hospitals, their performance meaningfully decreased for X-rays from new hospitals. By applying attribution methods to the classification decisions of the trained models, the authors were able to show that the models learned to accurately identify the hospital system that was used to acquire an X-ray, in combination with the specific department, allowing them to make accurate predictions on aggregate by simply learning the overall prevalence rates of these departments. Similar examples are imaginable in functional neuroimaging, as recently suggested by~\citet{chyzhyk_how_2022} who state that biomarker models for specific disease conditions could learn to distinguish patients from controls simply by their generally increased head motion.

For these reasons, we echo a recent call of machine learning researchers to always consider whether the application of complex models (such as DL models) is necessary to answer the research question at hand, or whether the application of simpler models, with better interpretability, could suffice~\citep{rudin_stop_2019}. While we do believe that DL models hold a high promise for mental state decoding research, e.g., with their ability to learn from large-scale neuroimaging datasets~\citep{schulz_performance_2022}, we also believe that many common mental state decoding analyses, which solely focus on few mental states in tens to a hundred of individuals, can be well-addressed with simpler decoding models with better interpretability~\cite[e.g.,][]{hoyos-idrobo_frem_2018,grosenick_interpretable_2013,kriegeskorte_information-based_2006,michel_total_2011,schulz_different_2020}.

We hope that with this work we can provide some insights into the strengths and weaknesses of prominent interpretation methods in mental state decoding, thereby enabling neuroimaging researchers to make an informed choice in situations where explanations for the mental state decoding decisions of DL models are needed.

%% file: sections/acknowledgments.tex
We gratefully acknowledge the support of NIH under No. U54EB020405 (Mobilize), NSF under Nos. CCF1763315 (Beyond Sparsity), CCF1563078 (Volume to Velocity), and 1937301 (RTML); ARL under No. W911NF-21-2-0251 (Interactive Human-AI Teaming); ONR under No. N000141712266 (Unifying Weak Supervision); ONR N00014-20-1-2480: Understanding and Applying Non-Euclidean Geometry in Machine Learning; N000142012275 (NEPTUNE); NXP, Xilinx, LETI-CEA, Intel, IBM, Microsoft, NEC, Toshiba, TSMC, ARM, Hitachi, BASF, Accenture, Ericsson, Qualcomm, Analog Devices, Google Cloud, Salesforce, Total, the HAI-GCP Cloud Credits for Research program, the Stanford Data Science Initiative (SDSI), the Texas Advanced Computing Center (TACC) at The University of Texas at Austin, and members of the Stanford DAWN project: Facebook, Google, and VMWare. The U.S. Government is authorized to reproduce and distribute reprints for Governmental purposes notwithstanding any copyright notation thereon. Any opinions, findings, and conclusions or recommendations expressed in this material are those of the authors and do not necessarily reflect the views, policies, or endorsements, either expressed or implied, of NIH, ONR, or the U.S. Government. 

FMRI data for the MOTOR and WM datasets were 
provided by the Human Connectome Project (HCP S1200 release), WU
Minn Consortium (Principal Investigators: David Van Essen and Kamil Ugurbil; 1U54MH091657)
funded by the 16 NIH Institutes and Centers that support the NIH Blueprint for Neuroscience
Research; and by the McDonnell Center for Systems Neuroscience at Washington University. FMRI data for the heat-rejection dataset were publicly 
shared in a preprocessed format by~\citet{kohoutova_toward_2020}.

%% file: supplements/methods/fmriprep.tex
\subsubsection{Fmriprep details}
\label{appendix_fmriprep_details}
Results included in this manuscript come from preprocessing performed
using \emph{fMRIPrep} 20.2.3 (\citet{fmriprep1}; \citet{fmriprep2};
RRID:SCR\_016216), which is based on \emph{Nipype} 1.6.1
(\citet{nipype1}; \citet{nipype2}; RRID:SCR\_002502).

\begin{description}
\item[Anatomical data preprocessing]
The T1-weighted (T1w) images were corrected for intensity
non-uniformity (INU) with \texttt{N4BiasFieldCorrection} \citep{n4},
distributed with ANTs 2.3.3 \citep[RRID:SCR\_004757]{ants}, and used as
T1w-reference throughout the workflow. The T1w-reference was then
skull-stripped with a \emph{Nipype} implementation of the
\texttt{antsBrainExtraction.sh} workflow (from ANTs), using OASIS30ANTs
as target template. Brain tissue segmentation of cerebrospinal fluid
(CSF), white-matter (WM) and gray-matter (GM) was performed on the
brain-extracted T1w using \texttt{fast} \citep[FSL 5.0.9,
RRID:SCR\_002823,][]{fsl_fast}. Volume-based spatial normalization to
two standard spaces (MNI152NLin2009cAsym, MNI152NLin6Asym) was performed
through nonlinear registration with \texttt{antsRegistration} (ANTs
2.3.3), using brain-extracted versions of both T1w reference and the T1w
template. The following templates were selected for spatial
normalization: \emph{ICBM 152 Nonlinear Asymmetrical template version
2009c} {[}\citet{mni152nlin2009casym}, RRID:SCR\_008796; TemplateFlow
ID: MNI152NLin2009cAsym{]}, \emph{FSL's MNI ICBM 152 non-linear 6th
Generation Asymmetric Average Brain Stereotaxic Registration Model}
{[}\citet{mni152nlin6asym}, RRID:SCR\_002823; TemplateFlow ID:
MNI152NLin6Asym{]},
\item[Functional data preprocessing]
For each of the BOLD runs found per subject, the following preprocessing was performed. First, a reference
volume and its skull-stripped version were generated using a custom
methodology of \emph{fMRIPrep}. Susceptibility distortion correction
(SDC) was omitted. The BOLD reference was then co-registered to the T1w
reference using \texttt{flirt} \citep[FSL 5.0.9,][]{flirt} with the
boundary-based registration \citep{bbr} cost-function. Co-registration
was configured with nine degrees of freedom to account for distortions
remaining in the BOLD reference. Head-motion parameters with respect to
the BOLD reference (transformation matrices, and six corresponding
rotation and translation parameters) are estimated before any
spatiotemporal filtering using \texttt{mcflirt} \citep[FSL
5.0.9,][]{mcflirt}. The BOLD time-series (including slice-timing
correction when applied) were resampled onto their original, native
space by applying the transforms to correct for head-motion. These
resampled BOLD time-series will be referred to as \emph{preprocessed
BOLD in original space}, or just \emph{preprocessed BOLD}. The BOLD
time-series were resampled into standard space, generating a
\emph{preprocessed BOLD run in MNI152NLin2009cAsym space}. First, a
reference volume and its skull-stripped version were generated using a
custom methodology of \emph{fMRIPrep}. Automatic removal of motion
artifacts using independent component analysis
\citep[ICA-AROMA,][]{aroma} was performed on the \emph{preprocessed BOLD
on MNI space} time-series after removal of non-steady state volumes and
spatial smoothing with an isotropic, Gaussian kernel of 6mm FWHM
(full-width half-maximum). Corresponding ``non-aggresively'' denoised
runs were produced after such smoothing. Additionally, the
``aggressive'' noise-regressors were collected and placed in the
corresponding confounds file. Several confounding time-series were
calculated based on the \emph{preprocessed BOLD}: framewise displacement
(FD), DVARS and three region-wise global signals. FD was computed using
two formulations following Power (absolute sum of relative motions,
\citet{power_fd_dvars}) and Jenkinson (relative root mean square
displacement between affines, \citet{mcflirt}). FD and DVARS are
calculated for each functional run, both using their implementations in
\emph{Nipype} \citep[following the definitions by][]{power_fd_dvars}.
The three global signals are extracted within the CSF, the WM, and the
whole-brain masks. Additionally, a set of physiological regressors were
extracted to allow for component-based noise correction
\citep[\emph{CompCor},][]{compcor}. Principal components are estimated
after high-pass filtering the \emph{preprocessed BOLD} time-series
(using a discrete cosine filter with 128s cut-off) for the two
\emph{CompCor} variants: temporal (tCompCor) and anatomical (aCompCor).
tCompCor components are then calculated from the top 2\% variable voxels
within the brain mask. For aCompCor, three probabilistic masks (CSF, WM
and combined CSF+WM) are generated in anatomical space. The
implementation differs from that of Behzadi et al.~in that instead of
eroding the masks by 2 pixels on BOLD space, the aCompCor masks are
subtracted a mask of pixels that likely contain a volume fraction of GM.
This mask is obtained by thresholding the corresponding partial volume
map at 0.05, and it ensures components are not extracted from voxels
containing a minimal fraction of GM. Finally, these masks are resampled
into BOLD space and binarized by thresholding at 0.99 (as in the
original implementation). Components are also calculated separately
within the WM and CSF masks. For each CompCor decomposition, the
\emph{k} components with the largest singular values are retained, such
that the retained components' time series are sufficient to explain 50
percent of variance across the nuisance mask (CSF, WM, combined, or
temporal). The remaining components are dropped from consideration. The
head-motion estimates calculated in the correction step were also placed
within the corresponding confounds file. The confound time series
derived from head motion estimates and global signals were expanded with
the inclusion of temporal derivatives and quadratic terms for each
\citep{confounds_satterthwaite_2013}. Frames that exceeded a threshold
of 0.5 mm FD or 1.5 standardised DVARS were annotated as motion
outliers. All resamplings can be performed with \emph{a single
interpolation step} by composing all the pertinent transformations
(i.e.~head-motion transform matrices, susceptibility distortion
correction when available, and co-registrations to anatomical and output
spaces). Gridded (volumetric) resamplings were performed using
\texttt{antsApplyTransforms} (ANTs), configured with Lanczos
interpolation to minimize the smoothing effects of other kernels
\citep{lanczos}. Non-gridded (surface) resamplings were performed using
\texttt{mri\_vol2surf} (FreeSurfer).
\end{description}

Many internal operations of \emph{fMRIPrep} use \emph{Nilearn} 0.6.2
\citep[\\RRID:SCR\_001362]{nilearn}, mostly within the functional
processing workflow. For more details of the pipeline, see
\href{https://fmriprep.readthedocs.io/en/latest/workflows.html}{the
section corresponding to workflows in \emph{fMRIPrep}'s documentation}.

\hypertarget{copyright-waiver}{%
\paragraph{Copyright Waiver}\label{copyright-waiver}}

The above boilerplate text was automatically generated by fMRIPrep with
the express intention that users should copy and paste this text into
their manuscripts \emph{unchanged}. It is released under the
\href{https://creativecommons.org/publicdomain/zero/1.0/}{CC0} license.

%% file: supplements/results/randomized-labels-performance.tex
\subsection{Data randomization test}

\begin{figure}[!h]
\begin{center}
\includegraphics[width=0.9\linewidth]{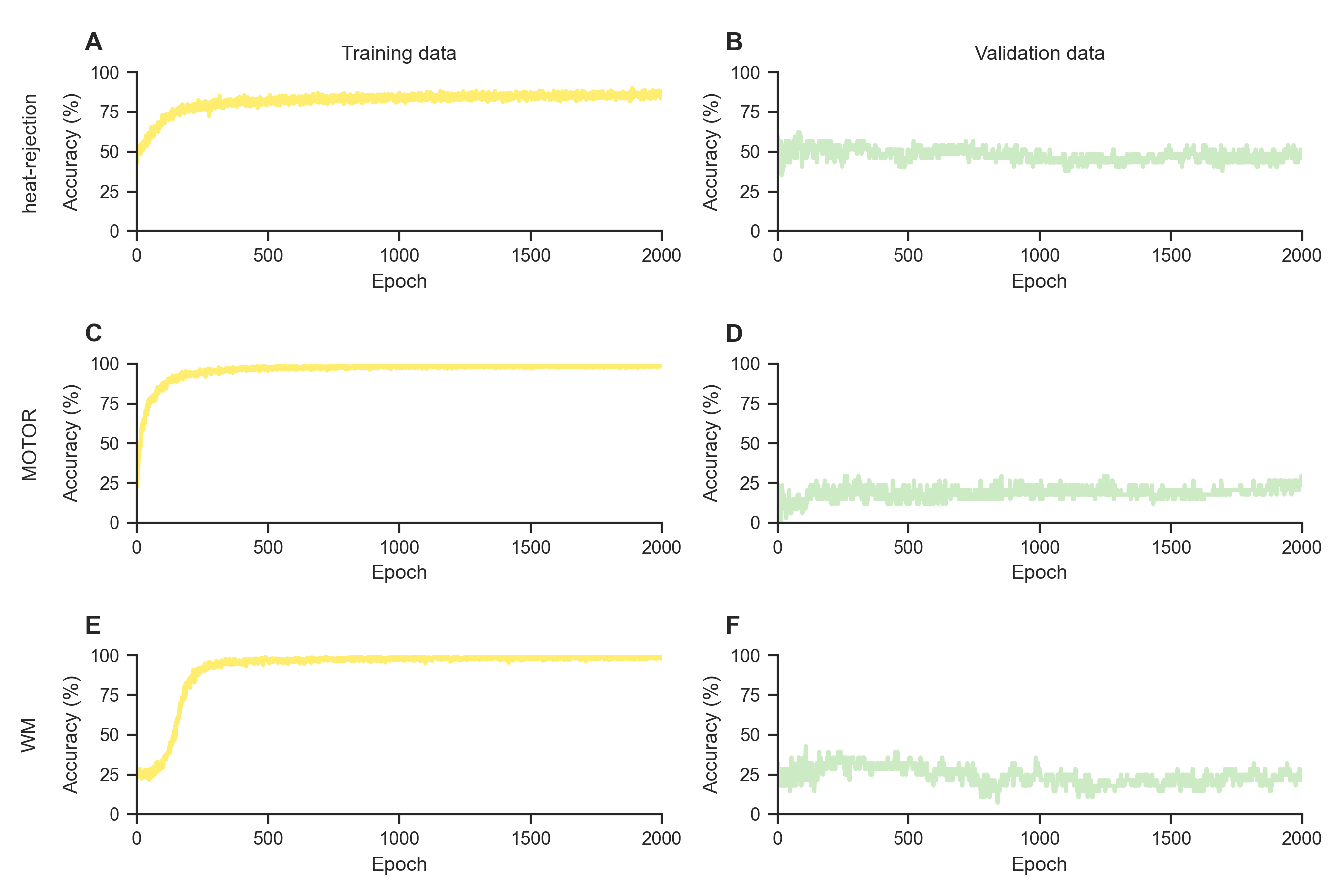}
\end{center}
\caption{Model performances in the data randomization test. We trained one DL model, according to each dataset's model configuration (see section \ref{results_hyperopt} of the main text), on a variant of each dataset with permuted mental state labels. A,C,E: All models learn to decode the randomized mental state labels well in the training data. B,D,F: Yet, the models perform close to chance in decoding the randomized labels from the validation data, indicating that the models simply learned to memorize the mapping between trial-level BOLD maps and mental state labels of the training data. As in our other analyses (see section \ref{results_decoding-performance} of the main text), we randomly separated each dataset's training data into new distinct training and validation datasets, comprising $95\%$ and $5\%$ of the data respectively. Colors indicate training (yellow) and validation (green) data. Lines indicate decoding accuracies over the course of training.}
\label{appendix:fig:results:randomized-labels}
\end{figure}